\documentclass[11pt,letter]{article}
\pdfoutput=1 

\usepackage{jheppub}
\usepackage{amssymb,amsmath,amsthm,graphicx}
\usepackage{graphics, color}
\usepackage{latexsym}
\usepackage{bm}
\usepackage[hang,flushmargin]{footmisc}
\usepackage{epsfig}
\usepackage{textcomp}
\usepackage{booktabs}
\usepackage{array}
\usepackage{caption}
\hypersetup{
 pdfauthor={William L. Boyce, Jorge  E. Santos},
 citecolor=black,linkcolor=black,urlcolor=black}
\usepackage[utf8]{inputenc}
\allowdisplaybreaks[1]

\def\ie{{\it i.e.} }
\def\eg{{\it e.g.} }

\begin{document}
\title{\centering{EFT Corrections to Charged \\ Black Hole Quasinormal Modes}}
\author{William~L.~Boyce}
\author{and Jorge~E.~Santos}

\affiliation{DAMTP, Centre for Mathematical Sciences, University of Cambridge, Wilberforce Road, \\ Cambridge CB3 0WA, UK}
\emailAdd{wlb26@cam.ac.uk}
\emailAdd{jss55@cam.ac.uk}

\newcommand{\blue}{\color{blue}}
\newcommand{\red}{\color{red}}

\abstract{We study the impact of higher-derivative corrections from Effective Field Theory on the quasinormal mode spectrum of Reissner–Nordström black holes. While previous work has explored corrections to Schwarzschild and Kerr black holes—typically using small-rotation approximations—a comprehensive analysis near extremality remains lacking. We focus on Reissner–Nordström black holes as a tractable model admitting an extremal limit, enabling investigation of the effect of these corrections on the so-called zero damped modes, which dominate in this regime. Specifically, we derive a corrected Moncrief equation governing quasinormal modes and present both analytic and numerical results for the corrected frequencies. This work also offers the first explicit test of the recently proposed “Quasinormal Mode Causality” bound \cite{melville2024causalityquasinormalmodesgreft}, which constrains Effective Field Theory coefficients by requiring that quasinormal mode lifetimes do not increase measurably under ultraviolet-complete Effective Field Theory corrections. Using Standard Model contributions—particularly those arising from integrating out the electron—we verify that this bound holds. Our results provide new insights into the interplay between Effective Field Theory, extremal black hole dynamics, and causality in gravitational theories.}

\maketitle

\section{Introduction}
\noindent\indent General relativity can be formulated as a low energy Effective Field Theory (EFT) of a larger, possibly UV complete, theory of physics. This EFT typically involves higher derivative corrections to the Einstein-Hilbert action, suppressed by powers of an energy scale at which new physics emerges \cite{Endlich_2017}. By studying these EFT corrections we can gain an insight into phenomenological effects of conjectured UV theories. We can also make connections to the Swampland program \cite{vafa2005stringlandscapeswampland}, which seeks to constrain EFT coefficients by appealing to necessary properties in the IR regime for a UV complete theory to exist.

Almost a decade after LIGO’s first successful detection of a binary black hole merger \cite{LIGOScientific:2016aoc}, it is hoped that quasinormal modes (QNMs) observed during the post-merger ringdown will become a key observable in gravitational wave astronomy in the coming years (see \cite{Berti:2025hly} for a comprehensive and recent review on QNMs). High-precision measurements of these modes could provide new insights into gravity beyond general relativity.

Recent progress has been made \cite{Cardoso_2018,Cano:2023tmv,Cano:2023jbk,Cano:2024bhh,Cano:2024jkd,Cano:2024wzo,cano2024higherderivativecorrectionskerrquasinormal,Maenaut:2024oci} in deriving the effects of EFT corrections on the QNMs of black holes. While the first work explores EFT corrections to the QNM spectrum of Schwarzschild black holes, the remaining body of work focuses on EFT corrections to the QNM spectrum of Kerr black holes. However, to make the problem tractable, approximation schemes were employed. In particular, all calculations were carried out in the small rotation limit, \ie using perturbation theory around Schwarzschild. As a result, we still lack an EFT calculation that focuses on black holes admitting an extremal limit and provides a comprehensive study of the full moduli space of solutions, including the approach to extremality—where perturbative schemes around the Schwarzschild black hole may break down. A notable exception is the study presented in \cite{Miguel:2023rzp}, which analysed EFT corrections to the QNM spectrum of Maxwell and scalar perturbations on a Kerr black hole background, with particular attention to the approach to extremality.

From an astrophysical perspective, EFT corrections to the Kerr black hole \cite{Kerr:1963ud} would be the most natural to study. However, this is a formidable task that we do not undertake here. Instead, in this paper, we consider EFT corrections to the Einstein-Maxwell action \cite{Kats:2006xp} and analyse their effect on the QNMs of a Reissner–Nordström (RN) black hole. Although charged black holes are not expected to exist in our universe, making this calculation observationally irrelevant, it represents the first investigation of EFT corrections to QNMs in the near-extremal regime. This regime is of particular interest due to recent works \cite{Horowitz_2023a,Horowitz_2023b,Horowitz_2024} showing that EFTs can break down near the horizon in the extremal limit.

This work is also of relevance to a recently conjectured `QNM Causality' bound on UV-complete theories, proposed in \cite{melville2024causalityquasinormalmodesgreft}. It is known that EFT corrections, caused by integrating out massive degrees of freedom, will push the singularities of linear response functions deeper into the complex plane - \ie if there is a pole at $\omega \in \mathbb C$, then integrating out a massive particle will cause $\left|\text{Im}(\omega)\right|$ to increase. It is argued that this property will then hold for QNM frequencies, which lie at the poles of Greens function obeying certain causal boundary conditions. Due to the finite resolving power of EFTs, it is only required that UV complete EFTs should cause no \emph{measurable} increase in the lifetime of QNMs relative to the uncorrected theory - this will result in bounds on Wilsonian coefficients in the EFT so that they obey this property. When viewed as the low-energy limit of the Standard Model, the EFT corrections to Einstein-Maxwell become calculable, with the leading contributions arising from integrating out the electron. We can substitute these corrections in and directly verify that this bound holds in this case.

The plan for this paper is as follows: We will begin in Sec. 2 with a review of the analytic methods to derive the Moncrief equation obeyed by QNMs of a RN black hole, including deriving some temperature scaling laws for frequencies whose waveforms remain supported near the horizon in the approach to extremality - \ie the Zero Damped Modes (ZDMs). In Sec. 3 we repeat the analysis of Sec. 2 in the EFT of Einstein-Maxwell, deriving an `EFT corrected Moncrief equation' and corrections to the ZDM frequencies. Sec. 4 will present some numerical calculations of the corrections to some frequencies, and verify numerically the EFT corrected scaling law for ZDMs derived in Sec. 3. We will conclude by verifying QNM Causality in this case by explicitly checking this with the EFT corrections induced by the electron on a RN black hole.

\section{Quasinormal Modes in Einstein-Maxwell Theory}
\subsection{The Moncrief Equation}
We begin with a brief review of the theory of QNMs of the RN Black Hole in Einstein-Maxwell theory, derived in \cite{PhysRevD.9.2707,PhysRevD.10.1057,Moncrief:1975sb, PhysRevD.9.860}. The action is
\begin{equation}
    S_{\text{EM}} = \int {\rm d}^4 x \sqrt{-g} \Bigg( \frac{1}{2 \kappa_4^2} R - \frac{1}{4} F_{ab}F^{ab} \Bigg)\,,
\end{equation}
where $F= {\rm d}A$. For convenience we work in units with $\kappa_4^2 = 2$. The metric and 1-form potential for the RN solution are given by\footnote{We do not consider magnetically charged black holes in this paper.}
\begin{subequations}
\begin{equation}
\begin{aligned}
    {\rm d}s_{\text{RN}}^2 &= -f(r) {\rm d}t^2 + \frac{{\rm d}r^2}{f(r)}+ r^2{\rm d}\Omega_2^2\\
    A_{\text{RN}} &= - \frac{Q}{r} {\rm d}t\,,
\end{aligned}
\end{equation}
where
\begin{equation}
f(r) = 1-\frac{2M}{r} + \frac{Q^2}{r^2}\,,
\label{eq:black}
\end{equation}
\label{RNmetric}%
\end{subequations}%
$M$ is the (Komar) mass, $Q$ is the electric charge, and ${\rm d}\Omega_2^2$ is the metric on a unit-radius round two-sphere, which we parametrize as
\begin{equation}
{\rm d}\Omega_2^2 = \frac{{\rm d}x^2}{1 - x^2} + (1 - x^2)\, {\rm d}\phi^2\,,
\end{equation}
with $x \in [-1, 1]$ and $\phi \sim \phi + 2\pi$. Here, $x$ is related to the standard polar coordinate on the two-sphere by $x \equiv \cos\theta$.

The metric is singular in $(t, r, x, \phi)$ coordinates when $f(r) = 0$ and when $r = 0$. While $r = 0$ corresponds to a curvature singularity, the same is not true for $f(r) = 0$, which represents a coordinate singularity. These coordinate singularities occur at
\begin{equation}
r = r_{\pm} = M \pm \sqrt{M^2 - Q^2}\,.
\end{equation}
For $M < |Q|$, $f(r)$ remains positive and the metric describes a naked singularity. From this point onward, we restrict to $M \geq |Q|$. 

To analytically continue the spacetime metric into the region $0 < r \leq r_+$, we introduce ingoing Eddington–Finkelstein coordinates,
\begin{equation}
{\rm d}v = {\rm d}t + {\rm d}r_{\star}
\end{equation}
where $r_{\star}$ is the Regge–Wheeler coordinate, defined by
\begin{equation}
{\rm d}r_* = \frac{{\rm d}r}{f(r)}\,.
\end{equation}
In these coordinates, the RN metric takes the form
\begin{equation}
{\rm d}s_{\text{RN}}^2 = -f(r) {\rm d}v^2 + 2\,{\rm d}v\,{\rm d}r + r^2 {\rm d}\Omega_2^2 \,.
\label{eq:efcoord}
\end{equation}
This metric is smooth across both $r = r_+$ and $r = r_-$, extending down into the region $r > 0$. The hypersurface $r = r_-$ is a Cauchy horizon, and the region $r_- \leq r \leq r_+$ lies inside a black hole region. In the $(v, r, x, \phi)$ coordinates, $r = r_+$ corresponds to the future event horizon of the black hole.

The horizon is Killing and has an associated surface gravity given by
\begin{equation}
\kappa = \frac{\sqrt{M^2 - Q^2}}{r_+^2}\,.
\end{equation}
When $r_+=r_-$, the horizon becomes degenerate, and is said to be extremal.

A QNM of this black hole is characterized by its frequency $\omega$ and angular momentum number $\ell$. Such a mode corresponds to a perturbation of the metric and the 1-form gauge potential of the form
\begin{equation}
{\rm d}s^2 = {\rm d}s^2_{\text{RN}} + h_{a b}{\rm d}x^{a}{\rm d}x^{b}\quad \text{and}\quad A = A_{\text{RN}} + \delta A\,.
\end{equation}
We will adopt the Regge-Wheeler gauge \cite{Regge:1957td}, in which case these perturbations are given by\footnote{Note that the inclusion of f(r) factors is different to standard conventions. These will be convenient when it comes to considering EFT perturbations.}:
\begin{equation}
\label{QNMmetric}
\begin{aligned}
    h_{ab} \, {\rm d}x^a  \,{\rm d}x^b =&e^{-i \omega t} \left\{ P_{\ell}(x) \left[H_0(r) \,{\rm d}t^2 + 2 i \omega \frac{H_1(r)}{f(r)} {\rm d}t \, {\rm d}r + \frac{H_2(r)}{f(r)^2} {\rm d}r^2 \right] \right.  \\
    &\left. +2 (1-x^2) P_{\ell}'(x) \left[h_0(r) {\rm d}t + i \omega \frac{h_1(r)}{f(r)} {\rm d}r\right] {\rm d}\phi + P_{\ell}(x) K(r) r^2{\rm d}\Omega_2^2 \right\} \\
    \delta A =& e^{-i \omega t}  \left\{ P_{\ell}(x)\left[k_1(r) {\rm d}t + i \omega \frac{k_2(r)}{f(r)}{\rm d}r\right] + (1-x^2)P_{\ell}'(x) k_3(r) {\rm d}\phi\right\}
\end{aligned}
\end{equation}
When the Einstein-Maxwell equations are satisfied, the functions $H_{0,1}, K, h_{0,1}, k_{1,2,3}$ can be expressed in terms of four master variables $Z_{1,2}^{\pm}(r)$ given by \cite{Chandrasekhar:1985kt}
\begin{subequations}
\begin{equation}
\begin{aligned}
    Z^{-}_1(r) &= \frac{2 q_1  k_3(r)}{\mu} + \frac{\sqrt{-q_1 q_2} h_1(r)}{r} \\
    Z^{-}_2(r) &= \frac{q_1 h_1(r)}{r} - \frac{2 \sqrt{-q_1 q_2}k_3(r)}{r}\\
    Z^{+}_1(r) &= \frac{\left(2 Q q_1+\mu  r \sqrt{-q_1
   q_2}\right)\left[r H_1(r) +r^2 K(r)\right] }{\mu  \left[r \left(6 M+\mu ^2 r\right)-4 Q^2\right]} + \frac{2 q_1 k_2(r)}{\mu}\\
   Z^{+}_2(r) &= \frac{\left(2 Q \sqrt{-q_1 q_2}-\mu  q_1 r \right)\left[r H_1(r) +r^2 K(r)\right] }{\mu  \left[r \left(6 M+\mu ^2 r\right)-4 Q^2\right]} + \frac{2 \sqrt{-q_1q_2} k_2(r)}{\mu}
\end{aligned}
\end{equation}
where we have defined the auxiliary variables
\begin{equation} \label{qdef}
\begin{aligned}
\mu &\equiv \sqrt{(\ell-1)(\ell+2)} \, , \\
    q_1 &\equiv 3 M + \sqrt{9 M^2 + 4 Q^2 \mu^2}\,,\\
    q_2 &\equiv 3 M - \sqrt{9 M^2 + 4 Q^2 \mu^2}\,.\\
\end{aligned}
\end{equation}
\end{subequations}%
An equivalent description of perturbations of spherically symmetric spacetimes can be given in terms of how a given perturbation transforms under the background $SO(3)$\footnote{This admits a generalisation to higher dimensions and is known as the Kodama–Ishibashi decomposition \cite{Ishibashi:2003ap,Kodama:2003jz,Kodama:2003kk}.}. Perturbations can then be built from harmonics on the unit-radius round two-sphere. On the $S^2$, there are no tensor harmonics, but scalar and vector harmonics do exist. The $(+)$ sector corresponds to scalar-derived gravito-electromagnetic perturbations, while the $(-)$ sector corresponds to vector-derived ones. These perturbations have different transformation properties under reflections across the equator of the $S^2$, with the $(+)$ sector transforming as $(-1)^{\ell}$ and the $(-)$ sector as $(-1)^{\ell+1}$.

Remarkably, the master variables then satisfy four \emph{decoupled} equations, known as the Moncrief equations
\begin{equation}\label{moncrief}
    f(r) \frac{{\rm d}}{{\rm d}r}\left[f(r) \frac{{\rm d}Z_{1,2}^{\pm}(r)}{{\rm d}r}\right] + \omega^2 Z_{1,2}^{\pm}(r) - f(r)\, V_{1,2}^{\pm}(r) \,Z_{1,2}^{\pm}(r) =0
\end{equation}
The Moncrief potentials $V_{1,2}^{\pm}(r)$ are given in terms of $M$ and $Q$ and are listed in \cite{Chandrasekhar:1985kt}.

To determine the QNM spectrum, we consider perturbations that are outgoing at future null infinity and regular across the future event horizon. The latter condition is enforced by requiring the perturbations to be regular in ingoing Eddington–Finkelstein coordinates near $r = r_+$, defined in the usual manner as $v=t+r_{\star}$ while the former requires regularity in outgoing null coordinates $u = t - r_{\star}$ as $r \to +\infty$.

The above considerations are enforced if
\begin{equation} 
    Z_{1,2}^{\pm}(r) \sim 
    \begin{cases} 
      \displaystyle e^{- i \omega r_*} \sim \left(1-\frac{r_+}{r}\right)^{-i \omega / 2 \kappa} & r_* \to - \infty, r \to r_+ \\
     \displaystyle  e^{i \omega r_*} \sim e^{i \omega r}\left(\frac{r}{r_+}\right)^{2 M i \omega } & r_*, r \to+\infty
   \end{cases}
\end{equation}

For numerical purposes, it will be convenient to change variables to
\begin{equation}
y = 1 - \frac{r_+}{r}
\end{equation}
so that the horizon is located at $y=0$, while asymptotic infinity is approached as $y \to 1^-$.

It is also convenient to define
\begin{subequations}
\begin{equation}
Z_{1,2}^{\pm}(r) = b(r)\, Y^{\pm}_{1,2}(r)
\end{equation}
with
\begin{equation}\label{boundaryfunction}
b(r) = \left(1 - \frac{r_+}{r}\right)^{-i \omega / 2 \kappa} \left(\frac{r}{r_+}\right)^{2 M i \omega} e^{i \omega r} = y^{-i \omega / 2 \kappa} (1 - y)^{-2 M i \omega} \exp\left(\frac{i \omega r_+}{1 - y}\right)
\end{equation}
\end{subequations}
so that $Y^{\pm}_{1,2}$ are now required to be regular at $y = 0$ and $y = 1$.

Eq. (\ref{moncrief}) is then equivalent to equations of the following form
\begin{equation} \label{regmoncrief}
    p_{1,2}^{\pm}(y) {Y_{1,2}^{\pm}}''(y)+q_{1,2}^{\pm}(y) {Y_{1,2}^{\pm}}'(y)+r_{1,2}^{\pm}(y) {Y_{1,2}^{\pm}}(y) =0
\end{equation}
where $p_{1,2}^{\pm}(y),q_{1,2}^{\pm}(y),r_{1,2}^{\pm}(y)$ are rational functions of $y$, which implicitly depend on $\omega$ and $M$. Imposing regularity of $Y^{\pm}_{1,2}$ at the horizon and at infinity picks out a discrete family of $\omega$s - these can be found by various numerical methods (see for instance \cite{Dias:2015nua}). This discrete set of frequencies is known as the quasinormal mode spectrum of Reissner–Nordström black holes and has been extensively studied in \cite{Chandrasekhar:1985kt}. These are the spectra we aim to study using EFT. While there are many interesting features in the QNM spectrum of RN black holes, we will focus particularly on the approach to extremality (when $M \to |Q|^+$). Two distinct behaviours appear in the QNM spectrum near extremality. Some modes, called zero-damped modes (ZDMs), have frequencies that vanish as extremality is approached, while others retain non-zero frequencies. The behaviour of ZDMs can be understood analytically, based solely on the geometry near the (almost) extremal horizon. We review this next.

\subsection{Zero Damped Modes} \label{ZDMsec}
We now summarise the behaviour of the ZDMs of a near extremal black hole, which can be understood as modes localised near the horizon \cite{Zimmerman_2016}. Recall the following derivation of the near-horizon geometry of the extremal RN \cite{Carroll_2009}, where we take both the extremal and near-horizon limits simultaneously. Begin with the RN metric in ingoing Eddington-Finkelstein coordinates as in Eq. (\ref{eq:efcoord}). Set $\epsilon = \sqrt{M^2 - Q^2}$, so that $r_{\pm} = M \pm \epsilon$. Define new coordinates $(y, \rho)$ by
\begin{equation}
    y =  \epsilon v/M^2 , \quad r = M + \epsilon \,\rho
\end{equation}
so that the metric becomes
\begin{equation}
    {\rm d}s_{\text{RN}}^2 = \frac{M^4}{(M + \epsilon \rho)^2} (1-\rho^2){\rm d}y^2+2 M^2  {\rm d}y \,{\rm d} \rho + (M  +\epsilon \rho)^2 {\rm d}\Omega_2^2 
\end{equation}
which limits to the near-horizon geometry AdS$_2 \times S^2$ in the $\epsilon \to 0$ limit. Now consider a mode in the exterior which remains supported in a neighbourhood of the horizon when $\epsilon \to 0$. Then, continuing this mode into the interior, the time dependence factor becomes $e^{- i \omega t} = e^{- i \omega v} e^{ i \omega r_*} = e^{-i M^2\omega   y/\epsilon}e^{ i \omega r_*}$. For this mode to survive in the this limit we must then have $\omega \sim \epsilon \sim \kappa$ as $\epsilon \to 0$. To account for this behaviour, it will be convenient to define $\omega = \kappa \tilde \omega/M$. Our task is to determine $\tilde \omega$.

Returning to the regularised Moncrief equation, we define a new coordinate $y = 2 \kappa z$. Then, to leading order in $\kappa$, Eq. (\ref{regmoncrief}) becomes
\begin{equation}\label{regNHmoncrief}
    \tilde{p}_{1,2}(z) {Y_{1,2}}''(z)+\tilde{q}_{1,2}(z) {Y_{1,2}}'(z)+\tilde{r}_{1,2}(z) {Y_{1,2}}(z) =0
\end{equation}
Note that we have dropped the $\pm$ superscripts - this is because both master variables obey the same equation in this limit. The coefficients in Eq. (\ref{regNHmoncrief}) are given by
\begin{equation}
\begin{aligned}
\tilde p_1 (z) &= \tilde p_2(z) = z(z+1)^2\\
\tilde q_1 (z) &= \tilde q_2(z) = -i (z+1) (\tilde \omega +(\tilde \omega+2 i) z+i)\\
\tilde r_1 (z) &= \frac{1}{4} \left[-4 \ell^2 (z+1)-12 \ell (z+1)-2 \left(\tilde \omega^2+i \tilde \omega+4\right)-\left(\tilde \omega^2+2 i \tilde \omega+8\right) z\right]\\
\tilde r_2(z) &=-\ell^2 (z+1)+\ell (z+1)-\frac{1}{4} \tilde \omega [2 (\tilde \omega+i)+(\tilde \omega+2 i) z] \\
\end{aligned}
\end{equation}
This equation can be readily solved in terms of Gauss hypergeometric functions
\begin{equation}\label{NHsolns}
\begin{aligned}
    Y(z) &= (1 +z)^{\sigma} \,{}_2F_1(\alpha,\beta;\gamma;-z) \\
    &= (1 +z)^{\sigma - \alpha}\, {}_2F_1\left(\alpha,\gamma-\beta;\gamma;\frac{z}{1+z}\right)\\
    &= (1+z)^{\sigma - \alpha}\frac{\Gamma(\gamma) \Gamma(\beta - \alpha)}{\Gamma(\beta) \Gamma(\gamma - \alpha)} {}_2F_1\left(\alpha,\gamma-\beta;\alpha - \beta +1;\frac{1}{1+z}\right)
    \\&+ (1+z)^{\sigma - \beta}\frac{\Gamma(\gamma) \Gamma(\alpha - \beta)}{\Gamma(\alpha) \Gamma(\gamma - \beta)} {}_2F_1\left(\gamma-\alpha,\beta;\beta - \alpha +1;\frac{1}{1+z}\right)
\end{aligned}
\end{equation}
where the first two lines are related by a Pfaff transformation, and the last two lines are a standard relationship between hypergeometric functions evaluated at $z$ and $1-z$ \cite{Whittaker_Watson_1996}. The parameters in this expression are given by
\begin{equation}\label{hypgeomcoeffsdefs}
\begin{aligned}
\alpha &= (-1)^g -\ell - i \tilde\omega  + \mathcal O(\kappa)\\
\beta &= 1 - (-1)^g + \ell - i \tilde\omega + \mathcal O(\kappa)\\
\gamma &= 1 - i \tilde\omega + \mathcal O(\kappa)\\
\sigma &= - i \tilde \omega/2 
\end{aligned}
\end{equation}
where $g = 1,2$ specifies whether we are working with the $Y_{1}$ modes or $Y_{2}$ modes. Note that for this equation to only have one regular solution near $z = 0$, we have assumed that $\gamma \notin - \mathbb{N}$.

To ensure that this mode remains localised within the horizon, we impose that $Y \to 0$ as $z \to \infty$. Since $\sigma - \alpha > 0$ for $\ell \geq 2$, our desired boundary condition is
\begin{equation}
Y(z) \sim z^{\sigma - \beta}
\end{equation}
so we must impose that either $\beta$ or $\gamma - \alpha$ lies at a pole of $\Gamma$. We can read off that $\gamma - \alpha > 0$, so we find the condition $\beta = -j$ for $j \in \mathbb{N}$. Hence, $\tilde{\omega} = -i (j + \ell + 1 - (-1)^g)$.

Note that we initially specified that $\gamma \notin -\mathbb{N}$, but this choice of $\tilde{\omega}$ would naively suggest otherwise. However, acknowledging the presence of the $\mathcal{O}(\kappa)$ corrections in the definitions in Eq.(\ref{hypgeomcoeffsdefs}) ensures that we can still sensibly assume this, as well as assume that the $\Gamma(\gamma)$ in the numerators in Eq.(\ref{NHsolns}) is not being evaluated at a pole (see \cite{Porfyriadis:2018yag} and \cite{Porfyriadis:2018jlw} for a more careful treatment, including the extension to the far region, which we neglected here).

\section{Quasinormal Modes in EFT Corrected Theories}
\subsection{EFT Corrections to the Reissner-Nordstr\"om Black Hole} \label{EFTintro}
Consider the EFT of Einstein-Maxwell theory as set out in \cite{Kats:2006xp}, which is given by
\begin{equation}\label{EFTaction}
\begin{aligned}
    S_{\text{EMEFT}} &= \int {\rm d}^4 x\,\sqrt{-g}\,\Bigg( \frac{1}{2 \kappa_4^2} R - \frac{1}{4} F_{ab}F^{ab}   + c_1 R^2 + c_2 R^{ab}R_{ab} + c_3 R_{abcd} R^{abcd} + c_4 R F_{ab} F^{ab}\\  & + c_5 R^{ab} {F_{a}}^{c} F_{bc} + c_6 R^{abcd} F_{ab} F_{cd}+ c_7 F_{ab} F^{ab} F_{cd} F^{cd} + c_8 F_{ab} F^{bc} F_{cd} F^{da}  \Bigg)\,.
\end{aligned}
\end{equation}
The equations of motion that follow from Eq. (\ref{EFTaction}) read
\begin{equation}
\begin{aligned}
\label{eqs:EOM}
\nabla^aF_{ab}&=c_4 J^{c_4}_{b}+c_5 J^{c_5}_{b}+c_6 J^{c_6}_{b}+c_7\,J^{c_7}_{b}+c_8 J^{c_8}_{b} \\ \\
R_{ab}-\frac{1}{2}Rg_{ab}-\kappa_4^2\left(F_{a}^{\phantom{a}c}F_{bc}{-}\frac{1}{4}g_{ab}F_{cd}F^{cd}\right) &=\kappa_4^2 \big(c_1 T^{c_1}_{ab}+c_2 T^{c_2}_{ab}+c_3 T^{c_3}_{ab}+c_4 T^{c_4}_{ab}
\\
&\qquad +c_5 T^{c_5}_{ab}+c_6 T^{c_6}_{ab}+c_7 T^{c_7}_{ab}+c_8 T^{c_8}_{ab}\big)\,,
\end{aligned}
\end{equation}
where
\begin{equation}
\begin{aligned}
J^{c_4}_a&=4\left(R\nabla^bF_{ba}-F_{ab}\nabla^b R\right)\\
J^{c_5}_a&=-2\left(R_{a}^{\phantom{a}c}\nabla_{b}F_c^{\phantom{c}b}+R^{cb}\nabla_b F_{ac}+F_{a}^{\phantom{a}c}\nabla_bR_{c}^{\phantom{c}b}+F^{cb}\nabla_{b}R_{ac}\right)\\
J^{c_6}_a&=-4 R_{adbc}\nabla^dF^{bc}-4 F^{bc}\nabla^d R_{adbc}\\
J^{c_7}_a&=8 \nabla^e\left(F_{ea} F^{cd}F_{cd}\right)\\
J^{c_8}_a&=-8 \nabla^b\left(F_{a}^{\phantom{a}p}F_{cp} F^{c}_{\phantom{c}b}\right)
\end{aligned}
\end{equation}
and
\begin{equation}
\begin{aligned}
T^{c_1}_{ab}& =4\nabla_a \nabla_b R-4 g_{ab}\Box R-4 R_{ab} R+g_{ab}R^2
\\
T^{c_2}_{ab}& =4\nabla_c \nabla_{(a} R_{b)}^{\phantom{b}c}-2\Box R_{ab}-2 g_{ab}\nabla_d\nabla_cR^{cd}-4 R_a^{\phantom{a}c}R_{bc}+g_{ab}R_{cd}R^{cd}
\\
T^{c_3}_{ab}& =-\left(4R_{a}^{\phantom{a}cde}R_{bcde}-g_{ab}R_{cdef}R^{cdef}+8 \nabla_c\nabla_d R_{(a\phantom{c}b)}^{\phantom{(a}c\phantom{b)}d}\right)
\\
T^{c_4}_{ab}& =4F^{cd}\nabla_{(a}\nabla_{b)}F_{cd}+4\nabla_a F^{cd}\nabla_b F_{cd}-4 g_{ab}F^{cd}\Box F_{cd}
\\
&\qquad-4 g_{ab}\nabla_e F_{cd}\nabla^e F^{cd}-2 R_{ab}F_{cd}F^{cd}-4 F_{a}^{\phantom{a}c}F_{bc}R+g_{ab}F_{cd}F^{cd}R
\\
T^{c_5}_{ab}& =4F_{(a}^{\phantom{(a}c}R_{b)d}F_{c}^{\phantom{c}d}-2F_{a}^{\phantom{a}c}F_{b}^{\phantom{b}d}R_{cd}+g_{ab}F_c^{\phantom{c}e}F^{cd}R_{de}-2\nabla_{(a}F_{b)}^{\phantom{b)}c}\nabla_d F_{c}^{\phantom{c}d}
\\
&\qquad-2\nabla_d \nabla_{(a}F_{b)c}F^{cd}-2\nabla_d\nabla_{(a}F^{cd}F_{b)c}-2 \Box F_{(a}^{\phantom{(a}c}F_{b)c}
\\
&\qquad -g_{ab}F^{cd}\nabla_d \nabla^e F_{ce}-2\nabla^d F_{(a}^{\phantom{(a}c}\nabla_{b)}F_{cd}-2 \nabla^d F_{a}^{\phantom{a}c}\nabla_d F_{bc}
\\
&\qquad+g_{ab}\nabla_c F^{cd}\nabla_e F_{d}^{\phantom{d}e}-g_{ab}F^{cd}\nabla_e\nabla_d F_{c}^{\phantom{c}e}-g_{ab}\nabla_d F_{ce}\nabla^e F^{cd}
\\
T^{c_6}_{ab}& =-\Big(6F_{(a}^{\phantom{bb}c}F^{de}R_{b)cde}-g_{ab}F^{cd}F^{ef}R_{cdef}-4F_{c(a}\nabla^{c}\nabla^{d}F_{b)d}
\\
&\qquad -4F_{d(a}\nabla^{d}\nabla^{c}F_{b)d}+4\nabla_{c}F_{a}^{\phantom{a}c}\nabla_dF_{b}^{\phantom{b}d}+4 \nabla_{c}F_{bd}\nabla^dF_{a}^{\phantom{a}c}\Big) \\
T^{c_7}_{ab}&=F^{pq}F_{pq}\left(g_{ab}F^{cd}F_{cd}-8 F_{a c}F_{b}^{\phantom{b}c}\right) \\
T^{c_8}_{ab}&=g_{ab}F_{c}^{\phantom{c}p}F_{dp}F^{cq}F^{d}_{\phantom{d}q}-8 F_{a}^{\phantom{a}p}F_{cp}F_{b}^{\phantom{b}q}F^c_{\phantom{c}q} \,.
\end{aligned}
\end{equation}
It is convenient to redefine the Wilsonian coefficients to have mass dimension $-2$, \ie
\begin{equation}
\begin{aligned}
    d_{1,2,3} &= \kappa_4^2 \,c_{1,2,3}\\
    d_{4,5,6} &= c_{4,5,6}\\
    d_{7,8} &= \kappa_4^{-2}\,c_{7,8}
\end{aligned}
\end{equation}
We are free to make a first order field redefinitions of the metric as in \cite{Cheung_2018},
\begin{equation}
    g_{ab} \mapsto g_{ab} + r_1 R_{ab} + r_2 R g_{ab} + r_3 \kappa_4^2 F_{ac} {F_b}^c + r_4 \kappa_4^2 g_{ab} F_{cd} F^{cd}\,,
\end{equation}
and we can also add a Gauss-Bonnet term to the action
\begin{equation}
   \frac{r_5}{\kappa_4^2}  \int {\rm d}^4x \sqrt{-g} \left( R^2 - 4 R_{ab} R^{ab} + R_{abcd} R^{abcd} \right)\,.
\end{equation}
Under these transformations no (classical) physically meaningful observables will be affected, but the Wilsonian coefficients shift by:
\begin{equation}\label{WilsonShifts}
\begin{aligned}
d_1 &\mapsto d_1 - \frac{1}{4} r_1 - \frac{1}{2} r_2 + r_5\\
d_2 &\mapsto d_2 + \frac{1}{2} r_1 - 4 r_5\\
d_3 &\mapsto d_3 + r_5\\
d_4 &\mapsto d_4 + \frac{1}{8} r_1 - \frac{1}{4} r_3 - \frac{1}{2} r_4\\
d_5 &\mapsto d_5 - \frac{1}{2} r_1 + \frac{1}{2} r_3\\
d_6 &\mapsto d_6\\
d_7 &\mapsto d_7 + \frac{1}{8} r_3\\
d_8 &\mapsto d_8 - \frac{1}{2} r_3\,.
\end{aligned}
\end{equation}
The above freedom allows us to set $c_{1,2,3,4,5} = 0$. Nevertheless, our analysis will hold for any choice of Wilsonian coefficients, so we can use the invariance of physical quantities under these redefinitions as a strong test of our analysis. In particular, we note that
\begin{equation}
d_0 \equiv d_2 + 4d_3 + d_5 + d_6 + 4d_7 + 2d_8\,, \quad d_6\,, \quad \text{and} \quad d_9 \equiv d_2 + 4d_3 + d_5 + 2d_6 + d_8
\end{equation}
are invariant under field redefinitions, and thus can serve as a convenient basis for expressing any classical observable in a field-redefinition-invariant form. For convenience, we again set $\kappa_4^2 = 2$.

We will be performing calculations to linear order in $c_K$, $K = 1,2, \ldots8$, so we can consider each perturbation separately. As such, in the following analysis we will drop the subscript and just work with some small EFT correction parametrised by $c$. We make an anzatz that the EFT corrected Reissner-Nordst\"om Black Hole has metric
\begin{subequations}
\begin{equation}
\begin{aligned}
    {\rm d}s_{\text{EFTRN}}^2 &= -f(r) {\rm d}t^2 + \frac{{\rm d}r^2}{g(r)}+ r^2{\rm d}\Omega_2^2\\
    A_{\text{EFTRN}} &= \psi(r) dt
\end{aligned}
\end{equation}
where
\begin{equation}
f(r) = f_0(r) +  c \, \delta f(r), \; g(r) = f_0(r) +  c \, \delta g(r), \;\psi(r) = - Q/r + c \,\delta\psi(r)
\end{equation}
\end{subequations}
with $f_0(r)$ given in Eq.~(\ref{eq:black}). We begin by perturbing the system in the microcanonical ensemble, meaning we require that the perturbations do not alter the mass or charge of the black hole. Consequently, the perturbations $\delta f$, $\delta g$, and $\delta \psi$ must decay sufficiently rapidly as $r \to \infty$. Under these conditions, it is straightforward to solve the Einstein-Maxwell equations to first order in perturbation theory and compute the corresponding functions, as listed in \cite{Kats:2006xp}.

To obtain a perturbation in the canonical ensemble, where the surface gravity (and thus the Hawking temperature) is held fixed, we adjust the mass via a shift $M \mapsto M + c\, \delta M$. The correction $\delta M$ is chosen such that it cancels the change in surface gravity $\kappa$ induced by the initial perturbation.\footnote{Such a choice of $\delta M$ is possible only when $|Q|/M \neq \sqrt{3}/2$, as the specific heat capacity $C = 2\pi\,\frac{\partial M}{\partial \kappa}$ diverges at that point.}

\subsection{EFT Corrections to the Moncrief Equation}
Given a choice of background, we now make a slight modification to the previous form of the metric and 1-form QNM perturbation
\begin{equation}
\begin{aligned}
    \delta h_{ab} \, {\rm d}x^a  \,{\rm d}x^b =&e^{-i \omega t} \left\{ P_{\ell}(x) \left[H_0(r) \,{\rm d}t^2 + 2 i \omega \frac{H_1(r)}{\sqrt{f(r)g(r)}} {\rm d}t \, {\rm d}r + \frac{H_2(r)}{f(r)g(r)} {\rm d}r^2 \right] \right.  \\
    &\left. +2 (1-x^2) P_{\ell}'(x) \left[h_0(r) {\rm d}t + i \omega \frac{h_1(r)}{\sqrt{f(r) g(r)}} {\rm d}r\right] {\rm d}\phi + P_{\ell}(x) K(r) r^2{\rm d}\Omega_2^2 \right\} \\
    \delta A =& e^{-i \omega t}  \left\{ P_\ell(x)\left[k_1(r) {\rm d}t + i \omega \frac{k_2(r)}{\sqrt{f(r)g(r)}}{\rm d}r\right] + (1-x^2)P_\ell'(x) k_3(r) {\rm d}\phi\right\}
\end{aligned}
\end{equation}
We define the auxiliary function
\begin{equation}
    \rho(r)= \mu^2 \frac{f(r)}{g(r)} + r \left[3 f'(r) + 2 r \left(\psi'(r)\right)^2 \right]\,,
\end{equation}
where $\mu$ and $q_{1,2}$ are as defined in Eq. (\ref{qdef}). We further define the master variables to be:
\begin{equation}\label{EFTmastervars}
\begin{aligned}
    Z^{-}_1(r) &= \frac{2 q_1  k_3(r)}{\mu} + \frac{\sqrt{-q_1 q_2} h_1(r)}{r}\,, \\
    Z^{-}_2(r) &= \frac{q_1 h_1(r)}{r} - \frac{2 \sqrt{-q_1 q_2}k_3(r)}{r}\,,\\
    Z^{+}_1(r) &= \frac{2 q_1  k_2(r)}{\mu} + \frac{(2 Q q_1 + \sqrt{-q_1 q_2} r \mu) [H_1(r) + r K(r)]}{r\mu\rho(r) }\,,\\
    Z^{+}_2(r) &= \frac{2 \sqrt{-q_1 q_2}  k_2(r)}{\mu} + \frac{(2 Q\sqrt{- q_1q_2}- q_1 r  \mu) [H_1(r) + r K(r)]}{r\mu\rho(r) }\,.
\end{aligned}
\end{equation}
The master variables, in turn, can be shown to satisfy an EFT-modified Moncrief equation (with the $\pm$, $1$, and $2$ labels suppressed for simplicity)\footnote{Here we are working with non-degenerate perturbation theory. Of course, the $+$ and $-$ modes are isospectral, so generically we would need degenerate perturbation theory for this analysis. However, in this paper we are only considering parity preserving EFT corrections, so there will be no mixing of these sectors.}
\begin{equation}
    \sqrt{f(r)g(r)}\frac{{\rm d}}{{\rm d}r}\left(\sqrt{f(r) g(r)} \frac{{\rm d} Z(r)}{{\rm d}r}\right) + \omega^2 Z(r) - \sqrt{f(r) g(r)} V(r) Z(r) = c \left[ \alpha (r) Z(r) + \beta(r) Z'(r) \right]
    \label{eq:central}
\end{equation}
where the $V(r)$ potentials are as in the unperturbed case, and the functions $\alpha(r)$ and $\beta(r)$ depend on $Q$, $M$, $\omega$ and $\ell$.

We convert Eq.~(\ref{eq:central}) to a regular form by considering the boundary function as defined in Eq. (\ref{boundaryfunction}), now using the EFT corrected $\kappa, r_+, M$ and $Q$. We further assume that the new modes frequencies are perturbatively far from an unperturbed frequency, i.e. $\omega = \omega_0 + c \, \delta \omega$. This gives the following equation
\begin{equation}
    p(y) Y''(y) + q(y) Y'(y) + r(y) Y(y) = c \Big[ \big(a_0(y) Y(y) + b_0(y) Y'(y)\big) + \delta \omega \big(a_\omega(y) Y(y)+ b_\omega(y) Y'(y)\big)\Big]
\end{equation}
where $p(y), q(y), r(y)$ are as defined for the unperturbed case i.e. with respect to $M_0, \omega_0$. Here $a_{0, \omega}(y)$ and $b_{0, \omega}(y)$ also depend on $M_0, \omega_0$, and $Q$. In what follows, we aim to determine $\delta \omega$ for several choices of the EFT coefficients, as a function of $Q$ and $M$.

We now assume that QNMs solutions to the EFT-modified Moncrief equation lie perturbatively close to the QNMs of the RN black hole. Specifically, we set
\begin{equation}
    Y(y) = Y_0(y) + c \, \zeta(y)
\end{equation}
where $Y_0$ obeys the unperturbed equation Eq. (\ref{regmoncrief}). We then expand to first order in the EFT coefficient $c$ and obtain an equation for $\zeta(y)$ and $\delta \omega$, which takes the following schematic form:
\begin{equation} \label{EFTregmoncrief}
    p(y) \zeta''(y) + q(y) \zeta'(y) + r(y) \zeta (y) =  a_0(y) Y_0(y) + b_0(y) Y_0'(y) + \delta \omega \left[a_{\omega}(y) Y_0(y) + b_{\omega}(y) Y_0'(y)\right]\,,
\end{equation}
We can always redefine $\zeta(y)$ by the homogeneous solution $\zeta(y) \mapsto \zeta(y) + Y_0(y)$. To remove this ambiguity we impose the boundary condition $\zeta(0) = 0$. It is important to note that $a_0, b_0, a_\omega, b_\omega$ are smooth functions of $y$, even at $y= 0,1$ - this is ensured by the careful construction of master variables in Eq. (\ref{EFTmastervars}) and EFT corrections to the boundary function in Eq. ($\ref{boundaryfunction}$).

\subsection{EFT Corrections to the Zero Damped Modes}
We now compute the EFT corrections to the ZDMs of a near extremal black hole in the Canonical Ensemble. Using the transformation of Sec (\ref{ZDMsec}), the EFT corrected regularised Moncrief equation becomes:
\begin{equation}\label{NHcorrection}
    \tilde p(z) \zeta''(z) + \tilde q(z) \zeta'(z) + \tilde r(z) \zeta(z) =  \tilde a_0(z) Y_0(z) + \tilde b_0(z) Y_0'(z) + \delta \tilde\omega \left(\tilde a_{\omega}(z) Y_0(z) + \tilde b_{\omega}(z) Y_0'(z)\right)
\end{equation}
for some $\tilde a_0, \tilde b_0, \tilde a_{\omega}$ and $ \tilde b_{\omega}$ that depend on $M_0, \tilde \omega_0$ also.

Consider the mode with $\tilde \omega_0 = - i(j + \ell + 1 - (-1)^{g})$ which is given by Eq. \eqref{NHsolns}. In the large $z$ limit, $Y_0(z) \sim z^{j + \sigma_0}$, with $\sigma_0 = - i \tilde\omega_0/2$. Substituting this into Eq. \eqref{NHcorrection} and taking this limit gives
\begin{equation}
    z^2 \zeta''(z) + \tilde q_0 z \zeta'(z) + \tilde r_0 \zeta(z) = (t_0 + t_{\omega} \delta \omega) z^{j + \sigma_0}
\end{equation}
where
\begin{equation}
\begin{aligned}
    q_0 &= 1 + (-1)^g-j-\ell \\
   r_0 &= \begin{cases}
        \displaystyle\frac{1}{4} (-8 + j^2 -10 \ell -3 \ell^2 +2 j \ell + 2 j) & g = 1\\
        \\
        \displaystyle\frac{1}{4} (j^2 +2\ell - 3 \ell^2 +2 j \ell - 2 j) & g = 2
       \end{cases}\\
    t_{\omega} &= -i \ell + \frac{i}{2} \left[2 (-1)^g -1\right]
\end{aligned}
\end{equation}
and $t_0$ is perturbation dependent. Since $z^{j+\sigma_0}$ solves the homogeneous part of this equation, we ansatz that the particular solution is $\zeta(z) = \zeta_0 \, z^{j+\sigma_0} \log(z)$. We can then calculate that
\begin{equation}\label{EFTcorrectedscaling}
\begin{aligned}
    \zeta_0 &= -\frac{t_0 + t_{\omega} \,\delta \tilde\omega}{1+2\ell -2(-1)^g}\\
    Y(z) &\sim z^{j + \sigma_0 + c \zeta_0}
\end{aligned}
\end{equation}

As in Sec. (\ref{ZDMsec}), we impose that this solution should keep the usual scaling rule of $Y(z) \sim z^{j + \sigma}$, where $\sigma = - i \tilde \omega/2$. Matching this with the exponent in Eq. \eqref{EFTcorrectedscaling} allows us to read off
\begin{equation}
\delta \tilde \omega = -i\frac{t_0}{1 + 2 \ell - 2 (-1)^g}\,.
\end{equation}

It is straightforward to extract $t_0$ from $a_0(y), b_0(y)$. In fact, for all perturbations we have considered $t_0$ is independent of $j$ giving the same correction to each overtone in a given $\ell$ sector. We then find that the EFT corrections to the ZDM labelled by $(\ell,j, g, \pm)$ generically break the degeneracy between the axial $(-)$ and polar $(+)$ sectors, and are given by
\begin{equation}\label{EFTZDM analytics}
\begin{aligned}
\omega_1^- &= - i \kappa \left[ \ell + j + 2 + (4 c_7 + c_8) \frac{8 \ell (\ell+1) (\ell+2)}{4 \ell^2+8 \ell+3} \right] + \mathcal O(\kappa^2)\\
\omega_1^+ &= - i \kappa \left( \ell + j + 2\right) + \mathcal O(\kappa^2)\\
\omega_2^- &= - i \kappa \left[ \ell + j + (4 c_7 + c_8) \frac{8 \ell \left(\ell^2-1\right)}{4 \ell^2-1}  \right] + \mathcal O(\kappa^2)\\
\omega_2^+ &= - i \kappa \left( \ell + j \right) + \mathcal O(\kappa^2)
\end{aligned}
\end{equation}
Note that we have not omitted the results for the $c_{1,2,3,4,5,6}$ corrections—it turns out that these contributions vanish. Although many of these corrections are zero at $\mathcal{O}(\kappa)$, numerical results demonstrate that there are still $\mathcal{O}(\kappa^2)$ corrections to these modes. It can also be seen that, since $4 c_7 + c_8 = \kappa_4^2 (4 d_7 + d_8) = \kappa_4^2(d_0 + d_6 - d_9)$ is invariant under the field redefinitions of Eq.~(\ref{WilsonShifts}), these expressions are indeed field redefinition invariant, as expected.

\section{Numerical Results}
\subsection{Damped Modes}\label{damped modes section}
We can solve the equations of the last two sections numerically using spectral methods \cite{Trefethen}. We solve Eq. (\ref{regmoncrief}) with $Y = Y_0,\; \omega = \omega_0$ by placing on a Chebyshev grid with $n = 300$ and working to a minimum precision of $100$ digits. At the horizon ($y=0$) we take 5 derivatives of Eq. (\ref{regmoncrief}) and impose this on the solution in the form of boundary conditions to ensure regularity at the horizon near extremality. At asymptotic infinity we impose the Robin conditions derived from Eq. (\ref{regmoncrief}) which impose a linear relation between $Y$ and its first order derivative. This procedure reframes the equation as a matrix problem, from which we can solve for $\omega_0$ using relaxation methods. Substituting this unperturbed solution into the RHS of Eq. (\ref{EFTregmoncrief}) turns this equation into a system of linear equations which can be solved for $\delta \omega$.

The invariance of $\delta \omega$ under the redefinitions of Eq. (\ref{WilsonShifts}) provides a helpful test that our analysis is correct, and indeed we find that this is invariant up to a tolerance for numerical errors. We can then restrict to considering just $c_{6,7,8}$ perturbations. As an additional test, we note that for $d_6 = 0$ and $d_8 = -2 d_7$, the EFT could originate from integrating out the axion \cite{Cheung_2014, PhysRevD.22.343, DUNNE_2005, Bastianelli_2009, Bastianelli_2012, heisenberg2006consequencesdiractheorypositron}. However, since the axion does not couple to scalar perturbations (\ie the (+) sector), we expect $\delta \omega = 0$ in this case. We have verified that this is indeed the case, up to numerical tolerance.\footnote{We would like to thank Maciej Kolanowski for suggesting this additional test.}

We obtain plots for the corrections to the modes with $\ell=2$ which in the Schwarzschild limit become the fundamental modes of the gravitational sectors, \ie $g = 2$. Plotting the $g=1$ modes, which we omit here, we see the same qualitative behaviour. These plots have been generated for both the microcanonical ensemble and the canonical ensemble. As a further test of the accuracy of our numerics, we compare our calculation of the $c_6$ correction to the $Z_1^{-}$ fundamental mode in the Schwarzschild case (which gives a non-trivial correction to the frequency) to the calculation of \cite{Miguel:2023rzp}. We find that our results agree with this calculation.

\begin{figure}[t!]
\centering
\includegraphics[width=\linewidth]{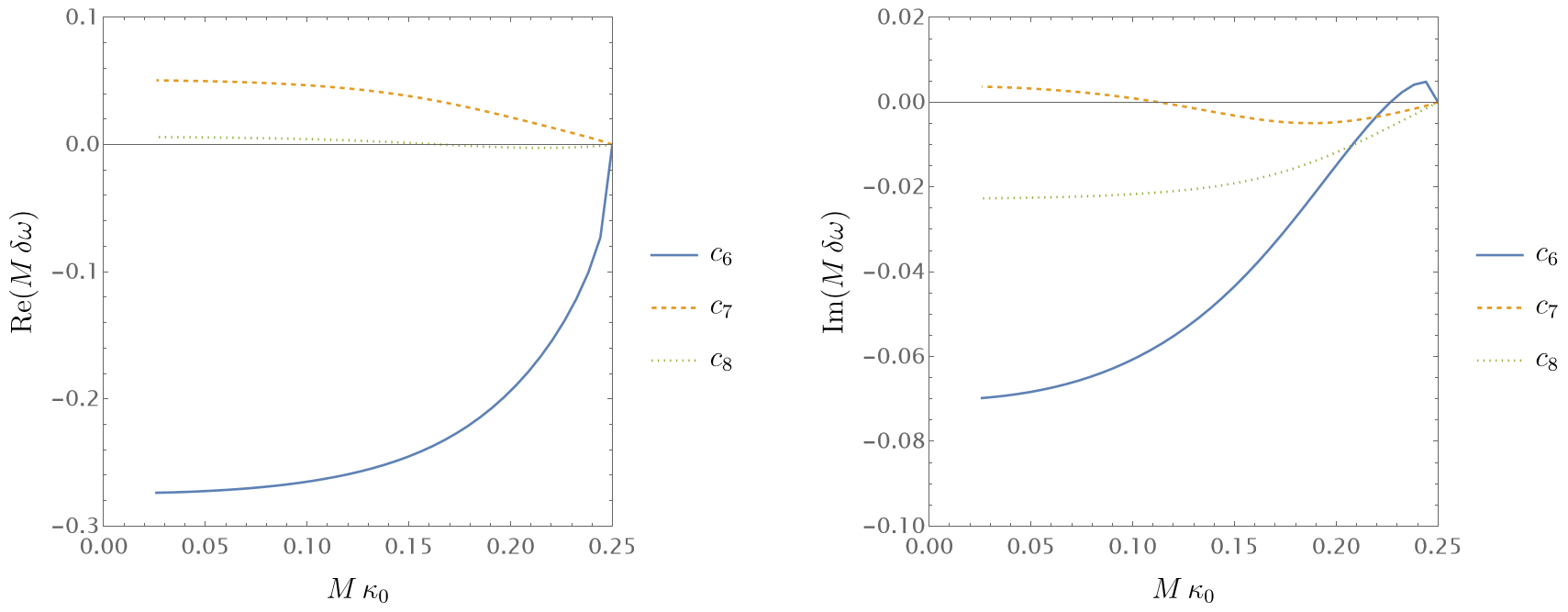}
    \caption{\label{fig:can1}EFT Corrections in the microcanonical ensemble, to the longest lived (\ie smallest $|\text{Im}(\omega_0 M_0)|$) \emph{damped}  mode in the $\ell=2$, $Z_2^- $ sector, as a function of $M \kappa_0$.}  
     \vspace{0.5cm}
    \includegraphics[width=\linewidth]{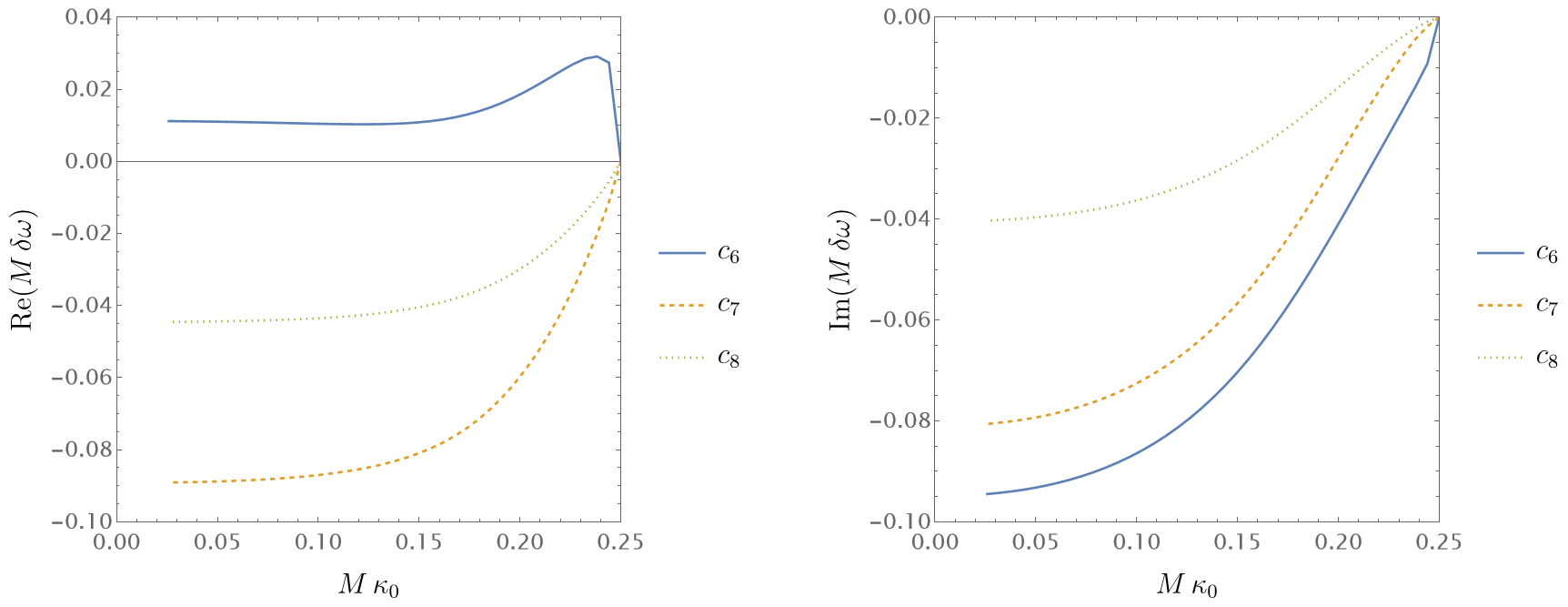}\\
    \caption{EFT Corrections in the microcanonical ensemble, to the longest lived (\ie smallest $|\text{Im}(\omega_0 M)|$, as a function of $M \kappa_0$.) \emph{damped} mode in the $\ell=2$, $Z_2^+ $ sector.}
\end{figure}

\begin{figure}[t!]
\centering
\includegraphics[width=\linewidth]{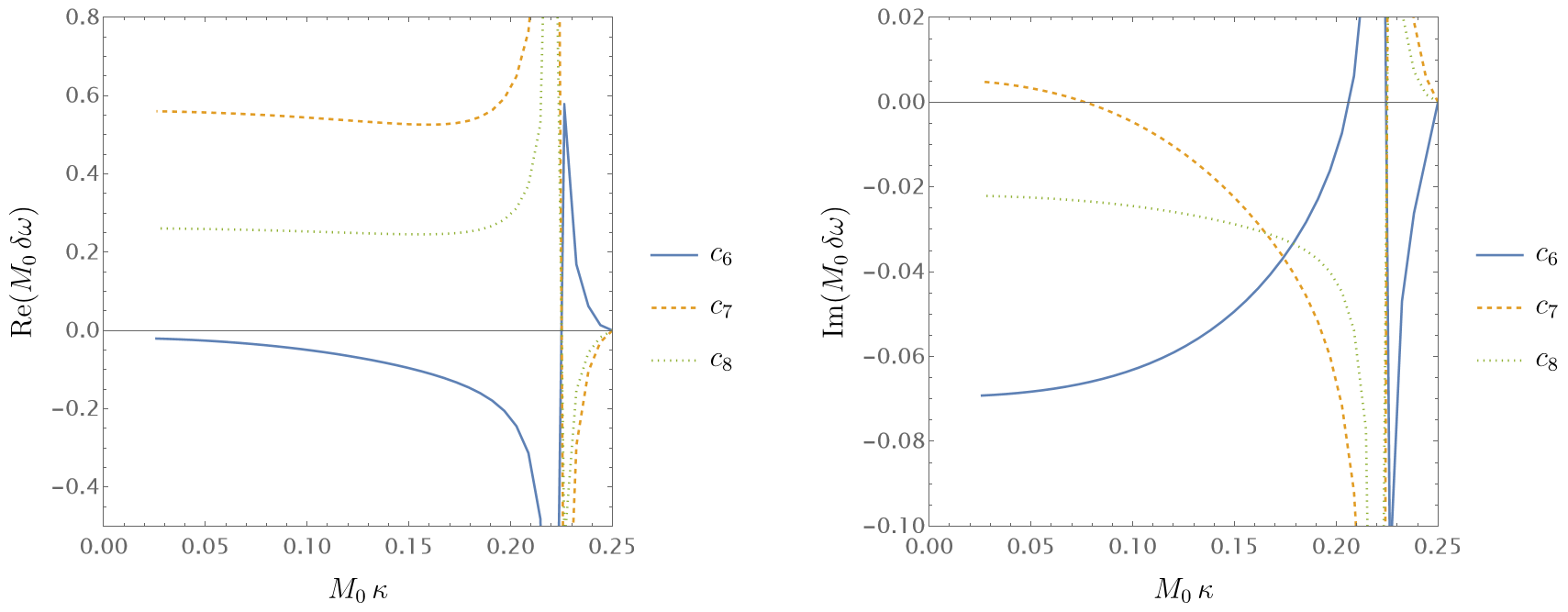}
    \caption{\label{fig:can2}EFT Corrections in the canonical ensemble, to the longest lived (\ie smallest $|\text{Im}(\omega_0 M_0)|$) \emph{damped}  mode in the $\ell=2$, $Z_2^- $ sector, as a function of $M_0 \kappa$.}  
    \vspace{0.5cm}
    \includegraphics[width=\linewidth]{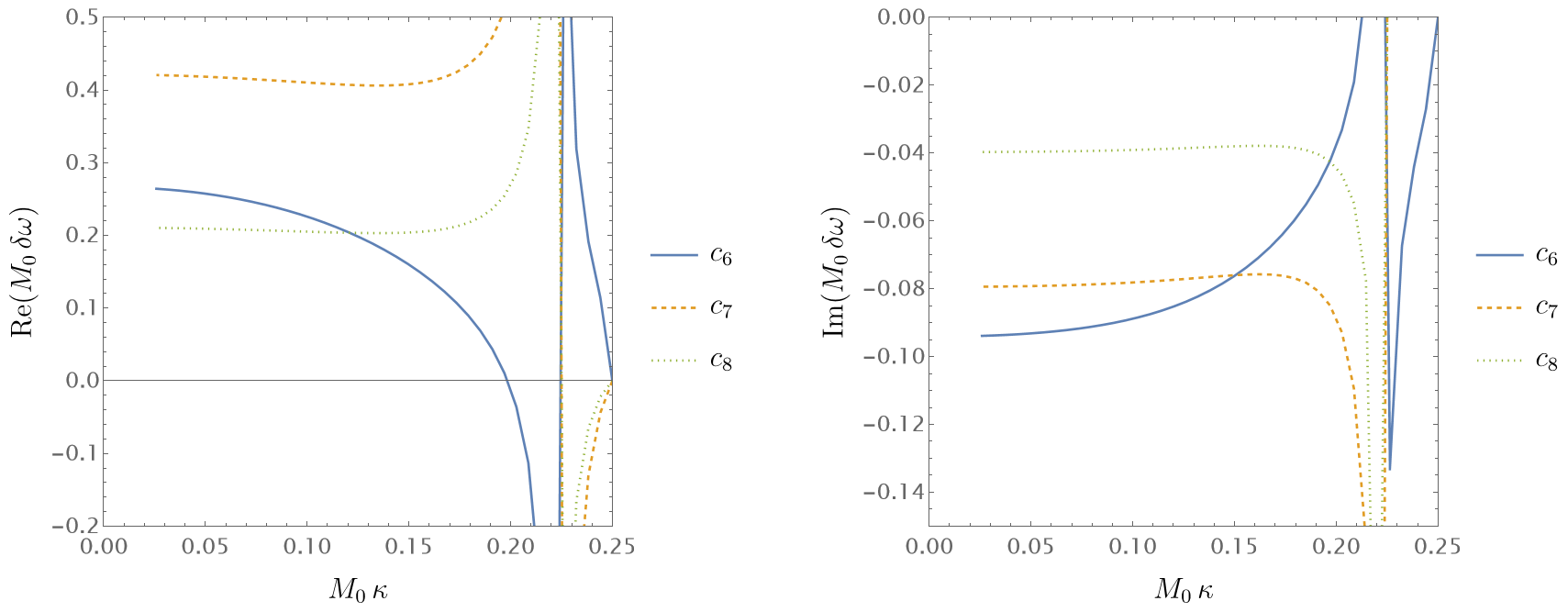}\\
    \caption{EFT Corrections in the canonical ensemble, to the longest lived (\ie smallest $|\text{Im}(\omega_0 M_0)|$) \emph{damped}  mode in the $\ell=2$, $Z_2^+ $ sector, as a function of $M_0 \kappa$.}
\end{figure}

\subsection{Zero Damped Modes}
Using the same methods as in Sec. \ref{damped modes section} we can calculate the EFTs correction to ZDMs numerically and verify agreement with the formulae Eq. (\ref{EFTZDM analytics}). As an example, we plot the corrections for $\delta \tilde \omega_2^-$ for $\ell=2, j=0$ due to $c_{6,7,8}$, with $n=250$ and to $300$ digits of precision.
\begin{figure}[t!]
\centering
    \includegraphics[width=9cm]{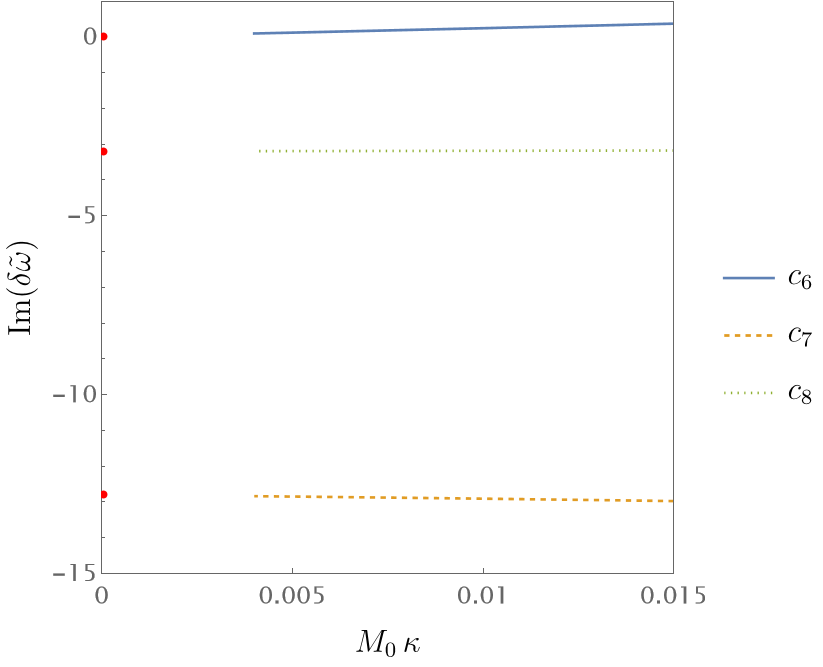}
    \caption{\label{fig:scaling}EFT Corrections in the canonical ensemble, to the lowest ZDM $\ell=2, g=2$ axial $(-)$ QNM, as a function of $M_0 \kappa$. The red discs give the analytical predictions for the $y$-intercepts of these lines.}
\end{figure}

A linear interpolation gives an estimate of the gradients of the graph, which can then be compared to the analytic prediction. We produce similar numerical plots for a range of choices of parameters and compare the gradients to analytic formulae.
\begin{table}[h]
\centering
\renewcommand{\arraystretch}{1.2}
\begin{tabular}{
    c c c c 
    >{\centering\arraybackslash}p{0.9cm} >{\centering\arraybackslash}p{0.9cm} >{\centering\arraybackslash}p{0.9cm}
    >{\centering\arraybackslash}p{0.9cm} >{\centering\arraybackslash}p{0.9cm} >{\centering\arraybackslash}p{0.9cm}
}
\toprule
 & & & & \multicolumn{6}{c}{$-\text{Im}(\delta \tilde \omega)$} \\
\cmidrule(lr){5-10}
$\ell$ & $g$ & $\pm$ & $j$ & \multicolumn{3}{c}{Numerical} & \multicolumn{3}{c}{Analytical} \\
 & & & & $c_6$ & $c_7$ & $c_8$ & $c_6$ & $c_7$ & $c_8$ \\
\midrule
2 & 1 & -- & 0 & 0.02 & 21.93 & 5.49 & 0 & 21.94 & 5.49 \\
2 & 1 & -- & 1 & 0.09 & 22.52 & 5.71 & 0 & 21.94 & 5.49 \\
2 & 2 & -- & 0 & 0.00 & 12.79 & 3.20 & 0 & 12.80 & 3.20 \\
2 & 2 & -- & 1 & 0.00 & 12.79 & 3.20 & 0 & 12.80 & 3.20 \\
2 & 2 & -- & 2 & 0.01 & 12.84 & 3.23 & 0 & 12.80 & 3.20 \\
2 & 1 & + & 0 & 0.02 & 0.02 & 0.01 & 0 & 0 & 0 \\
2 & 2 & + & 0 & 0.01 & 0.00 & 0.00 & 0 & 0 & 0 \\
2 & 2 & + & 1 & 0.01 & -0.01 & 0.00 & 0 & 0 & 0 \\
3 & 1 & -- & 0 & 0.04 & 30.57 & 7.66 & 0 & 30.48 & 7.62 \\
3 & 2 & -- & 0 & 0.01 & 21.93 & 5.49 & 0 & 21.94 & 5.49 \\
3 & 2 & -- & 1 & 0.01 & 21.97 & 5.50 & 0 & 21.94 & 5.49 \\
3 & 2 & + & 0 & 0.01 & 0.01 & 0.00 & 0 & 0 & 0 \\
4 & 2 & -- & 0 & 0.01 & 30.46 & 7.62 & 0 & 30.48 & 7.62 \\
\bottomrule
\end{tabular}
\caption{Comparison of numerical and analytical values for $-\text{Im}(\delta \tilde \omega)$}
\end{table}

These gradients are estimated near $\kappa = 0.01$. As shown in Fig.~\ref{fig:scaling}, $\delta \omega/\kappa$ has not yet reached a fully linear regime at this scale, so some inaccuracy in the estimation is expected. The table shows that the analytic and numerical results typically differ in the second decimal place, which we attribute to this estimation error.

\subsection{A Test of QNM Causality}
In \cite{melville2024causalityquasinormalmodesgreft}, several conjectures were proposed concerning the behaviour of QNMs under EFT corrections. Melville observed that, since integrating out heavy fields pushes the poles of linear response functions deeper into the complex plane—\ie increasing $\left|\text{Im}(\omega)\right|$—a similar effect should apply to QNM frequencies, which correspond to poles of causal Green’s functions. This leads to a proposed bound: EFT corrections should not produce a measurable increase in QNM lifetimes relative to the uncorrected theory, thereby constraining Wilson coefficients. In this work, we test this causality-based bound.

We can use our data to compute the relevant coefficient quantifying this effect, as defined in \cite{melville2024causalityquasinormalmodesgreft}
\begin{equation}
    \rho = \delta \tau \; \left|\text{Re} (\omega_0)\right| = \frac{\text{Im}(\delta \omega) \left|\text{Re}(\omega_0)\right|}{\text{Im}(\omega_0)^2} = \frac{1}{\kappa_4^2}\left[\rho_0 \,d_0 + \rho_6 \, d_6 + \rho_9\, d_9\right]\,,
    \label{eq:causa}
\end{equation}
\begin{figure}[t!]
\centering
    \includegraphics[width=7.5cm]{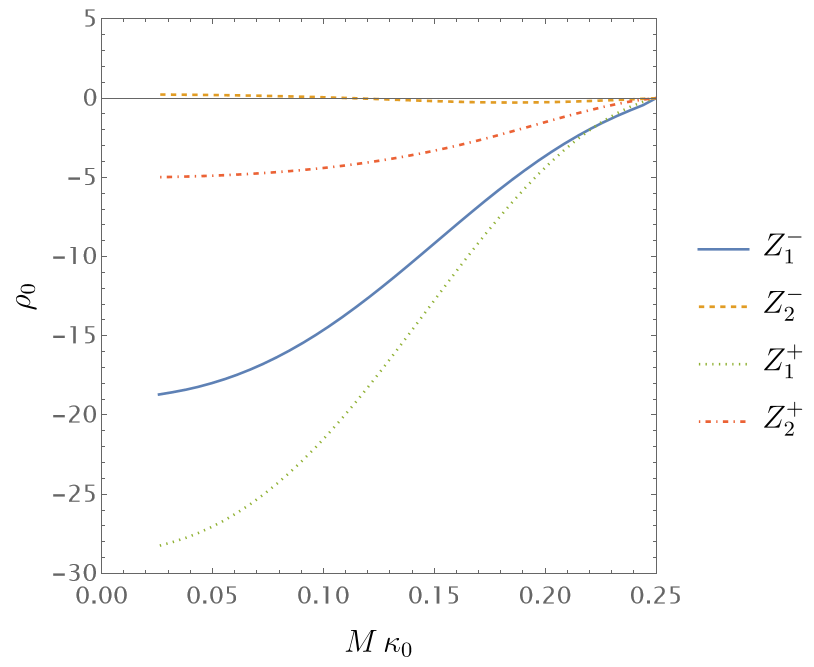}
    \includegraphics[width=7.5cm]{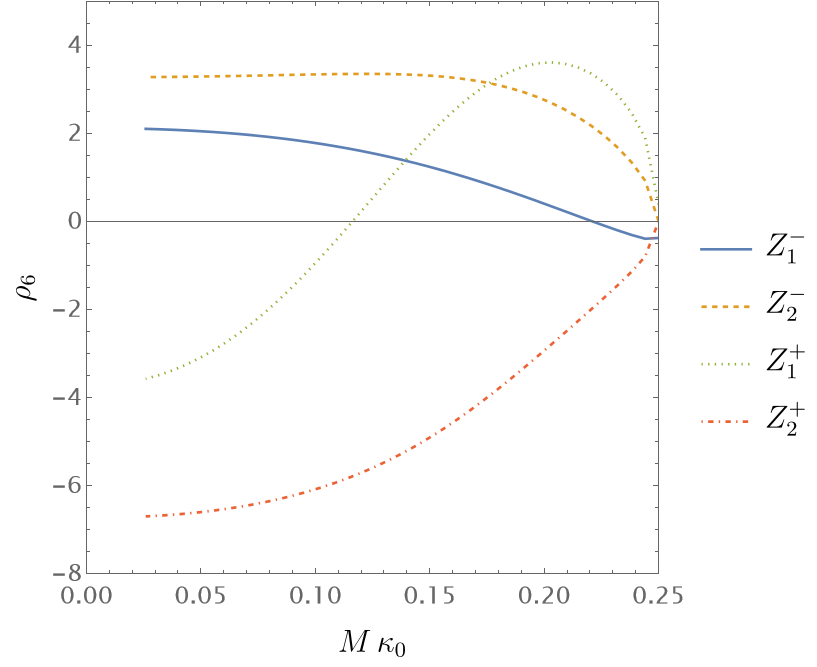}
    \includegraphics[width=7.5cm]{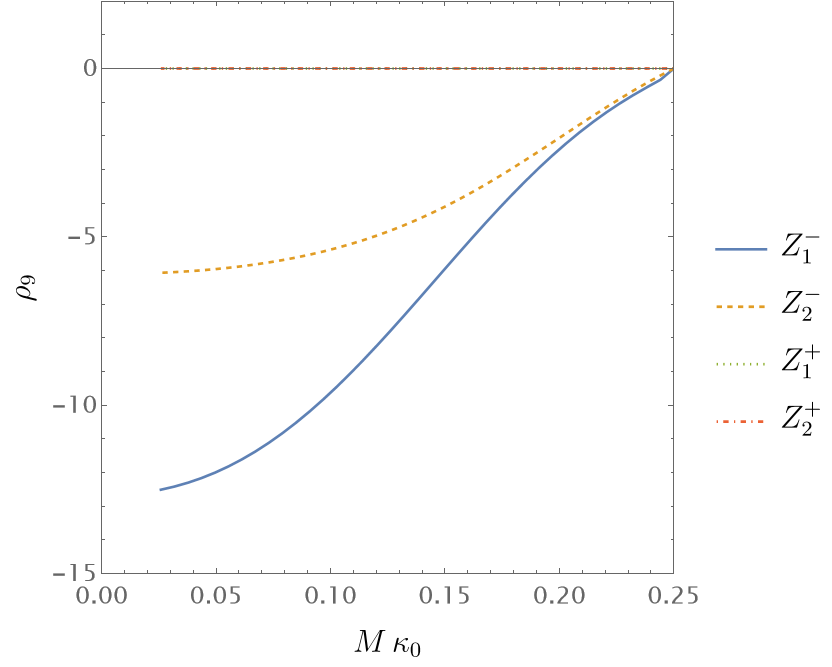}
    \caption{\label{fig:qnmbound}  We plot $\rho_{0,6,9}$, defined in Eq.~(\ref{eq:causa}), over a range of temperatures. This is done for the longest lived damped mode in the relevant sectors.}
\end{figure}
where $\rho_{0,6,9}$ are the relevant dimensionless contributions for each of the field redefinition invariants, which we plot in Fig.~\ref{fig:qnmbound} for the fundamental modes in the microcanonical ensemble\footnote{If we were to apply this test in the canonical ensemble, then the divergence at $|Q|/M = \sqrt{3}/2$ - see Figs.~\ref{fig:can1}-\ref{fig:can2} - would certainly violate this bound. This does not seem to be `in the spirit' of how this bound should be applied.} over the full range of $\kappa$. In \cite{melville2024causalityquasinormalmodesgreft}, this quantity is conjectured to be bounded above by $\rho \lesssim 1$. Note that for ZDMs, $\rho$ vanishes identically, since these modes are purely imaginary and therefore trivially satisfy the bound. Note that $\rho_{0,9}\lessapprox0$ whereas for $\rho_6$ this sign can vary depending on $\kappa$ and the choice of mode. Given the formulae in \cite{Cheung_2014, PhysRevD.22.343, DUNNE_2005, Bastianelli_2009, Bastianelli_2012, heisenberg2006consequencesdiractheorypositron}, it can be verified that if the coefficients $d_0$, $d_6$, and $d_9$ are generated by integrating out the electron (or indeed \emph{any} charged scalar or fermion, provided that they have mass $m \ll m_{\text{Pl}} \equiv \sqrt{8 \pi}/\kappa_4$), then $\rho < 0$ for all values of $\kappa$. Note that, in principle, there exists an infinite number of inequalities to check, corresponding to all possible overtones within each $(\pm, g)$ sector. We find it likely that the strongest constraints arise from the lowest-lying modes with ${\rm Re}(M \omega_0) \neq 0$, i.e., EFT corrections to the fundamental damped modes. If it had been the case that $\rho>0$ for \eg a charged fermionic field, then this would have given some upper bound on the charge-to-mass ratio. The absence of such a bound here is in contrast to bounds provided by unitarity and microcausality, which can constrain the charge-to-mass ratios of states in the theory\footnote{We would like to thank Scott Melville for pointing this out.} \cite{Hamada_2019, Bellazzini_2019,Arkani_Hamed_2022}.

\section{Conclusion}

In this work, we have explored the impact of higher-derivative corrections from Effective Field Theory on the quasinormal mode spectrum of RN black holes, focusing especially on the near-extremal regime where zero damped modes dominate. By deriving and analysing an EFT-corrected Moncrief equation, we obtained analytic and numerical results for the corrected quasinormal frequencies, providing a more complete picture beyond approximated methods used in previous studies.

Our work was partially motivated by \cite{Horowitz_2023a,Horowitz_2023b,Horowitz_2024}, where it was shown that EFT breaks down for near-extremal black holes due to the presence of tidal force-type singularities. For a RN black hole, the deformations causing such singularities can be classified as scalar-derived and vector-derived deformations, based on their transformation properties under rotations of the background $SO(3)$. In \cite{Horowitz_2023a}, scalar deformations were found not to produce these singularities, but more recently, \cite{Horowitz_2024} demonstrated that vector deformations lead to infinite tidal forces near extremal horizons. This raises the question of whether such effects leave imprints on classical observables, such as the QNM spectrum of RN black holes. Our results suggest that such signatures are unlikely, as we observed no significant differences in $\delta \omega$ between scalar-derived and vector-derived QNMs. However, the effect identified in \cite{Horowitz_2024} was significantly stronger when considering EFT corrections to Kerr-Newman black holes, which may imply that studying EFT corrections to their quasinormal mode spectrum is a promising direction for detecting such observables. That said, the quasinormal mode spectrum of Kerr-Newman black holes was only recently well understood \cite{Dias:2015wqa,Carullo:2021oxn,Dias:2021yju,Dias:2022oqm} (see also \cite{Davey:2023fin} for a comprehensive study of scalar QNMs in Kerr–Newman black holes), meaning that performing an EFT analysis in this context is currently beyond the reach of existing techniques. Alternatively, one could investigate EFT corrections to the QNM spectrum of charged black holes in Anti-de Sitter spacetime, where both scalar- and vector-derived deformations have been shown to produce significant tidal effects \cite{Horowitz_2023b}.

Importantly, we have tested the recently proposed Quasinormal Mode Causality bound, which places constraints on EFT coefficients by demanding no measurable increase in quasinormal mode lifetimes in UV-complete theories. Using calculable corrections from the Standard Model—specifically those arising from integrating out the electron—we verified that this causality bound is satisfied in our setup.

Our results contribute to understanding how EFT corrections influence black hole perturbations near extremality and highlight the subtle interplay between low-energy effective descriptions, black hole physics, and fundamental causality requirements. This work lays the groundwork for future investigations of more astrophysically relevant rotating black holes and the potential observational signatures in gravitational wave data.

\section*{Acknowledgements}
We are grateful to Maciej~Kolanowski, Scott~Melville, Harvey~Reall, and Grant~Remmen for their comments on an earlier draft. WLB was supported by an STFC studentship ST/Y509127/1. The work of JES was partially supported by STFC consolidated grant ST/X000664/1 and by Hughes Hall College.

\bibliography{references.bib}{}
\bibliographystyle{utphys-modified}

\end{document}